\documentclass[epj]{svjour}
\usepackage{graphics}
\usepackage{amsmath}
\usepackage{braket}
\usepackage[toc,page]{appendix}
\usepackage{mathtools}
\usepackage{extarrows}
\usepackage{tensor}
\usepackage{cancel}

\usepackage{ucs}
\usepackage[utf8]{inputenc}
\newcommand{\be}{\begin{equation}}
\newcommand{\ee}{\end{equation}}
\newcommand{\bea}{\begin{eqnarray}}
\newcommand{\eea}{\end{eqnarray}}

\begin{document}
\title{Are we survivors of the sudden past singularity?}

\titlerunning{Are we survivors of the sudden past singularity?}


\author{Adam Balcerzak\inst{1,2} \and Tomasz Denkiewicz\inst{1,2} \and Mateusz Lisaj\inst{3}
}

\authorrunning{A. Balcerzak, T. Denkiewicz, M. Lisaj}
%
%
\institute{Institute of Physics, University of Szczecin,
Wielkopolska 15, 70-451 Szczecin,  Poland \and Copernicus Center for Interdisciplinary Studies, Szczepa\'nska 1/5, 31-011 Krak\'ow, Poland \and  
Institute of Mathematics, Physics and Chemistry, Maritime University of Szczecin, Wa{\l }y Chrobrego 1-2, 70-500 Szczecin, Poland}
\date{Received: date / Revised version: date}
%
\abstract{In this paper, we investigate the viability of cosmological models featuring a type II singularity that occurs during the past evolution of the Universe. We construct a scenario in which the singularity arises and then constrain the model parameters using observational data from Type Ia Supernovae, Cosmic Chronometers, and Gamma Ray Bursts. We find that the resulting cosmological models based on scenarios with the past type II singularity cannot be excluded by kinematical tests using current observations. 
} 
\maketitle

\section{Introduction}
\label{sec:sec0}

Tippler's criterion \cite{tipler} distinguishes between weak and strong spacetime singularities based on whether structures can withstand the tidal forces within them. The application of this definition to cosmology has led to the discovery of interesting singular cosmological scenarios. Besides obviously strong cosmological singularities like the Big Bang and the Big Rip, weak singularities have been uncovered in various scenarios \cite{barrow,barrow2,nojiri,denkiewicz}. These scenarios have distinct behavior of the scale factor, the pressure and the energy density at the time of the singularity. It is worth noting that the evolution of the universe can be extended beyond any weak singularity \cite{lazkoz,lazkoz2}.

The Sudden Future Singularity (SFS), commonly referred to as a Type II singularity, emerges in a homogeneous universe described by the FLRW metric \cite{barrow}. These singularities have also been identified in anisotropic cosmologies \cite{barrow2}. This study focuses on constructing scenarios centered around SFS without relying on a specific equation of state. Instead, the scale factor is defined as a unique function of time. The singularity manifests at a finite time point, characterized by a divergence in pressure. However, both the scale factor and energy density remain constant. Notably, the SFS only breaches the dominant energy condition. Recent studies have proposed SFS as potential contributors to dark energy \cite{hoda,denkiewicz2}.

Historically, Type II singularities in cosmological models have been instrumental in illustrating the scale-de\-pen\-dent evolution of perturbations in dark matter and dark energy \cite{denkiewicz3}. These singularities were predominantly believed to arise after the current phase of the Universe's evolution. However, their mild nature, indicating potential evolution post-occurrence \cite{adam_t}, makes it plausible to introduce cosmological scenarios with past Type II singularities and assess their feasibility against observational data.

A promising research avenue involves identifying past signatures of such singularities. This investigative direction is being actively pursued. For instance, in \cite{Odintsov:2022eqm}, the authors delved into the implications of a late-time pressure singularity on the Earth's and solar system's historical trajectory. They theorized that the elliptical orbits of planets around the sun and the moon around the Earth might have been subtly influenced by this pressure singularity. Such alterations could lead to observable shifts in the Earth's climate and sea levels. Additionally, \cite{Odintsov:2022umu} explored the potential impact of a past pressure singularity on the shadows and photon orbits of cosmological supermassive black holes, suggesting that these effects could be discernible for future generations.

Moreover, the impact of other weak singularities, such as Type IV, on subsequent cosmic evolution has been previously investigated. Within the $F(R)$ gravity framework, the concept of Type IV singular bouncing has been analyzed and compared with other bouncing cosmological models \cite{Odintsov:2015zza,Odintsov:2015ynk}. This exploration extended to the evolution of inflationary paradigms, emphasizing both inflaton fields and modified gravity theories predicting sudden future singularities, especially Type IV singularities. In the broader context of general scalar-tensor cosmologies and models steered by phenomenological equations of state, research has delved into the influence of singularities on cosmological primordial perturbations and Hubble flow parameters \cite{Odintsov:2016plw,Odintsov:2015ynk,Odintsov:2015gba,Odintsov:2015jca,Nojiri:2015wsa,Nojiri:2015fia,Nojiri:2015fra}.

Our paper is organized as follows. In Sec. \ref{sec:sec1}, we present a scenario featuring a type II singularity that occurs during the past evolution of the Universe. Next, in Sec. \ref{sec:name2}, we provide details about the data used to constrain the parameters of the model based on the aforementioned scenario. Finally, in Sec. \ref{sec:Results}, we present the results of our analysis and draw conclusions about the viability of the proposed model.

\section{Type II singularity scenario} 
\label{sec:sec1}

As we have previously mentioned type II singularities may appear in homogeneous and isotropic cosmological framework governed by the Friedmann equations:
\bea \label{rho} \varrho(t) &=& \frac{3}{8\pi G}
\left(\frac{\dot{a}^2}{a^2} + \frac{kc^2}{a^2}
\right)~,\\
\label{p} p(t) &=& - \frac{c^2}{8\pi G} \left(2 \frac{\ddot{a}}{a} + \frac{\dot{a}^2}{a^2} + \frac{kc^2}{a^2} \right)~,
\eea
where $a\equiv a(t)$ is the scale factor, the dot represents a differentiation with respect to time $t$, $\rho(t)$ is the energy density, $p(t)$ denotes the pressure, $G$ and $c$ represent the gravitational constant and the speed of light, respectively, and $k=0, \pm 1$ is the curvature index. The Bianchi identity in such a framework is:
\be
\label{conser}
\dot{\varrho}(t) = - 3 \frac{\dot{a}}{a}
\left(\varrho(t) + \frac{p(t)}{c^2} \right)~.
\ee
A cosmological scenario featuring a type II singularity in the past can be described by the following scale factor:
\be 
\label{sf2} a(y) = a_s \left[\delta + \left(1 - \delta \right) y^m - \delta  |1 - y| ^n \right]~, \hspace{0.5cm} y \equiv \frac{t}{t_s} ,
\ee
where $t_s$ is the time of occurrence of the type II singularity and $a_s,\delta, m, n$ are constants with the contraint $1<n<2$.

The difference between the type II singularity scenario considered in \cite{barrow} and the scenario investigated in this paper is that the latter includes properly defined evolution for times that come after the singularity. Let us notice that in the limit $\delta\rightarrow0$ one retrieves the standard Friedmann limit with no singularity at all. Additionally, we will assume that $\delta<0$ since only models within that range exhibit accelerated expansion.

The scale factor (\ref{sf2}) in the limit $t\rightarrow0$ (close to the Big-Bang) scales as $y^m$ which emulates a barotropic perfect fluid with a barotropic index $w=-1 + 2/3m$ and the standard characteristics of early Universe cosmology do not change provided an appropriate value of $m$ is assumed. Take also note that condition $w\geq-1/3$, in which case none of the energy conditions is violated, translates into $0<m\leq1$ in the limit $t\rightarrow0$.

The redshift $z$ of a cosmological object in the considered model is given by:
\be \label{zet}
1+z = \frac{a(y_0)}{a(y_1)} = \frac{\left[\delta + \left(1 - \delta \right) y_0^m -
\delta  |1 - y_0| ^n \right]~}{\left[\delta + \left(1 - \delta \right) y_1^m -
\delta  |1 - y_1| ^n \right]~},
\ee
where $y_0 \equiv t_0/t_s$ and $y_1 \equiv t_1/t_s$  with $t_0$ and $t_1$ being the times of signal reception and emission, respectively. 

The dimensionless luminosity distance for a cosmological model described by  FLRW flat ($k=0$) metric is: 
\be \label{lumdist}
d_L=(1+z)\int_{0}^{z}\frac{dz'}{E(z')},
\ee
with $E(z)=H(z)/H_0$ where $H(z)$ denotes a Hubble parameter and $H_0$ represents its present-day value. Since we are bound to use the explicite form of the scale factor due to the lack of an analytic form of the equation of state the most appropriate form of (\ref{lumdist}) is:
\be \label{lumdist2}
d_L(z,\boldsymbol{p})= \left(1+z\right)a'(y_0)\int_{y_1}^{y_0}\frac{dy}{a\left(y\right)},
\ee
where $\boldsymbol{p}  = (\delta,m,n,y_0)$ is  is a vector composed of the parameters of the model. Note that the lower limit of integration $y_1$ in (\ref{lumdist2}) can be computed using (\ref{zet}) for a specified values of the parameters $\delta$, $m$, $n$, $y_0$ and $z$.
\\

\section{Observational constraints}

\label{sec:name2}
\subsection{Type Ia Supernovae}
The Pantheon dataset \cite{Scolnic:2017caz} comprises 1048 Type Ia supernovae (SNeIa) distributed over the redshift range $0.01<z<2.26$. The corresponding $\chi^2_{SN}$ statistic is computed as
\begin{equation}
\chi^2_{SN} = \Delta \boldsymbol{\mathcal{\mu}}^{SN} \; \cdot \; \mathbf{C}^{-1}_{SN} \; \cdot \; \Delta  \boldsymbol{\mathcal{\mu}}^{SN} \;,
\end{equation}
where $\Delta\boldsymbol{\mathcal{\mu}} = \mathcal{\mu}_{\rm theo} - \mathcal{\mu}_{\rm obs}$ is the difference between the theoretical and observed values of the distance modulus for each SNeIa, and $\mathbf{C}_{SN}$ represents the total covariance matrix. It should be noted that the full dataset is used, rather than a binned version (as in Anagnostopoulos et al. 2020). The distance modulus is defined as
\begin{equation}
\mu(z,\boldsymbol{p}) = 5 \log_{10} [ d_{L}(z, \boldsymbol{p}) ] +\mu_0 .
\end{equation} 
Due to the degeneracy between the Hubble constant $H_0$ and the SNeIa absolute magnitude (both of which are included in the nuisance parameter $\mu_0$), we marginalize $\chi^{2}_{SN}$ over $\mu_0$ obtaining \cite{conley}
\begin{equation}\label{eq:chis}
\chi^2_{SN}=a+\log \left(\frac{d}{2\pi}\right)-\frac{b^2}{d},
\end{equation}
where $a\equiv\left(\Delta \boldsymbol{\mathcal{\mu}}_{SN}\right)^T \; \cdot \; \mathbf{C}^{-1}_{SN} \; \cdot \; \Delta  \boldsymbol{\mathcal{\mu}}_{SN}$, $b\equiv\left(\Delta \boldsymbol{\mathcal{\mu}}^{SN}\right)^T \; \cdot \; \mathbf{C}^{-1}_{SN} \; \cdot \; \boldsymbol{1}$, $d\equiv\boldsymbol{1}\; \cdot \; \mathbf{C}^{-1}_{SN} \; \cdot \;\boldsymbol{1}$ and $\boldsymbol{1}$ is the identity matrix.

\subsection{Cosmic Chronometers}
The definition of CC is used for Early-Type galaxies which exhibit a passive evolution and a characteristic feature in their spectra \cite{Jimenez:2001gg}, for which can be used as clocks and provide measurements of the Hubble parameter $H(z)$ \cite{Moresco:2010wh}. The sample we are going to use in this work is from \cite{moresco} and covers the redshift range $0<z<1.97$. The $\chi^2_{H}$ is defined as
\begin{equation}\label{eq:hubble_data}
\chi^2_{H}= \sum_{i=1}^{24} \frac{\left( H(z_{i},\boldsymbol{p})-H_{obs}(z_{i}) \right)^{2}}{\sigma^2_{H}(z_{i})} \; ,
\end{equation}
where $\sigma_{H}(z_{i})$ are the observational errors on the measured values $H_{obs}(z_{i})$.

\subsection{Gamma Ray Bursts}
The standardization of Gamma Ray Bursts (GRBs) is still a subject of debate. However, we focus on the "Mayflower" sample consisting of 79 GRBs in the redshift range $1.44<z<8.1$ \cite{Liu:2014vda}, as it has been calibrated using a robust cosmological model-independent procedure. The distance modulus is the observable probe related to GRBs, and the same method used for SNeIa is applied here. The goodness-of-fit measure for GRBs, $\chi^2_{GRB}$, is given by \\ $\chi^2_{GRB}=a+\log d/(2\pi)-b^2/d$ as well, with $a\equiv \left(\Delta\boldsymbol{\mathcal{\mu}}^{G}\right)^T \, \cdot \, \mathbf{C}^{-1}_{G} \, \cdot \, \Delta  \boldsymbol{\mathcal{\mu}}^{G}$, $b\equiv\left(\Delta \boldsymbol{\mathcal{\mu}}^{G}\right)^T \, \cdot \, \mathbf{C}^{-1}_{G} \, \cdot \, \boldsymbol{1}$ and $d\equiv\boldsymbol{1}\, \cdot \, \mathbf{C}^{-1}_{G} \, \cdot \, \boldsymbol{1}$.

\section{Results and conclusions}
\label{sec:Results}
\begin{figure*}[p]
    \centering
    \includegraphics[width=\textwidth]{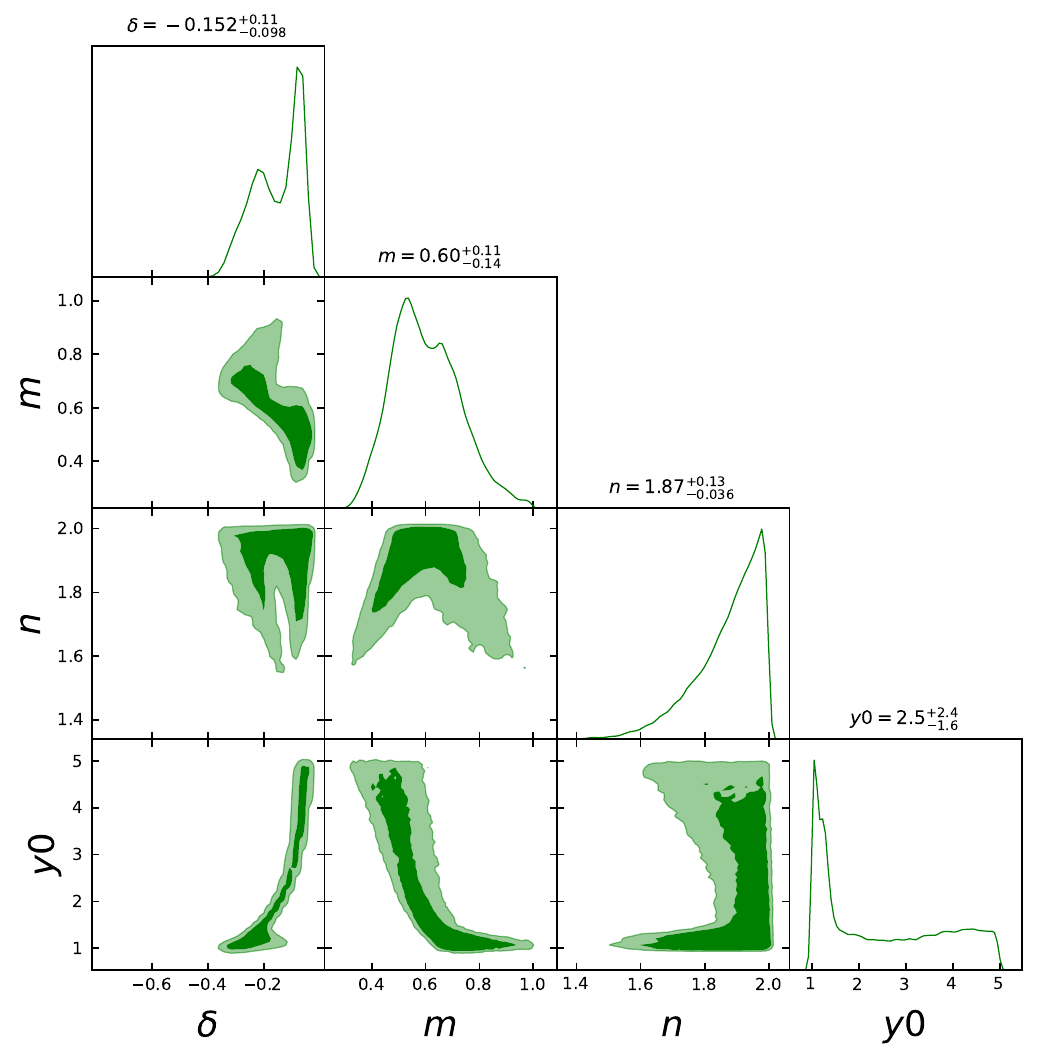}
    \caption{The contour plots display the marginalized distributions for model parameter pairs, computed jointly for SN Ia, CC, and GRBs. Two confidence intervals, corresponding to 68\% and 95\% credibility levels (indicated by dark green and light green areas, respectively), are shown. The distributions for each parameter of the model are also presented.}
    \label{ct}
\end{figure*}

We employed the MCMC Metropolis Hastings method to perform the analysis of the cosmological models featuring a type II singularity in the past evolution of the Universe. The total $\chi^2$ value used in the analysis was
\be\label{totchi}
\chi^2=\chi^2_{SN}+\chi^2_{H}+\chi^2_{GRB}.
\ee
The resulting contour plots in Fig. \ref{ct} illustrate the marginalized distributions of model parameter pairs, computed jointly for Type Ia Supernovae, Cosmic Chronometers, and Gamma Ray Bursts (the distributions for each parameter of the model are also displayed). Within these contour plots, two confidence intervals corresponding to the 68\% and 95\% credibility levels are indicated by dark green and light green areas, respectively. Our analysis reveals that the resulting cosmological models based on scenarios with the past type II singularity cannot be excluded by the kinematical tests using the current observational data. We can see that there exists an admissible value of $m=2/3$, which aligns with the characteristics of a dust-filled Einstein-de-Sitter universe in the  vicinity of Big Bang singularity.
 The marginalized distribution for $y_0$ attains its highest value at the leftmost boundary of its prior aligning with the present moment in the evolution of the Universe. The fact that the distribution is highest at the left edge, indicating a higher likelihood or probability density for $y_0<1$, suggests that the data tends to favor cosmological scenarios where the singularity occurs in the future (the proper Sudden Future Singularities). The validity of such models has been established in \cite{denkiewicz2} , where it was demonstrated that these models successfully satisfy numerous observational constraints. 

In conclusion main result of our work is that we demonstrated the viability of cosmological models featuring a type II singularity in the past evolution of the Universe. The presented plots provide evidence that these models remain consistent with the current observational data. Having sudden singularities in the past was already considered in the earlier works \cite{Odintsov:2022umu,Odintsov:2015zza,Odintsov:2022eqm,Odintsov:2016plw,Odintsov:2015ynk,Odintsov:2015gba,Odintsov:2015jca,Nojiri:2015wsa,Nojiri:2015fia,Nojiri:2015fra}, and in principle such a scenario can not be simply excluded. It would be interesting to further search of signatures of occurrence of such type of singularities in the past. For example in the observed large scale structure of the universe, which would result from the occurrence of the singularity in the early stages of the evolution during the period of forming cosmic background radiation or taking place directly in the times of structure formation after the time of recombination.  Or signatures in the cosmic background radiation itself. However, it is worth noting that the data appears to prefer models featuring the type II singularity in the future.

\end{document}